\documentclass[
10pt,aps,
notitlepage, 
bibnotes,
prb,
twocolumn,
showpacs,preprintnumbers,superscriptaddress,
amsmath,amssymb,floatfix]{revtex4-1}

\usepackage{graphicx}
\usepackage{dsfont}
\usepackage{dcolumn}
\usepackage{bm}
\usepackage{color}
\usepackage{xcolor}
\usepackage{url}
\usepackage{tabularx}
\usepackage{array}
\usepackage{booktabs}
\usepackage{comment}
\usepackage{hyperref}
\usepackage{footnote}
\usepackage{lipsum}
\usepackage[normalem]{ulem}
\usepackage{booktabs}

\definecolor{colorA}{rgb}{0, 0.5, 0.5}
\definecolor{colorB}{rgb}{0.5, 0, 0.9}
\definecolor{colorC}{rgb}{0.4, 0, 0.4}
\definecolor{color_green}{rgb}{0, 0.39, 0}

\makeatletter
  \def\my@tag@font{\normalsize}
  \def\maketag@@@#1{\hbox{\m@th\normalfont\my@tag@font#1}}
  \let\amsmath@eqref\eqref
  \renewcommand\eqref[1]{{\let\my@tag@font\relax\amsmath@eqref{#1}}}
\makeatother

\begin{document}
\title{Tailed skyrmions -- an obscure branch of magnetic solitons}

\author{Vladyslav~M.~Kuchkin}
\email{vkuchkin@hi.is}
\affiliation{Science Institute, University of Iceland, 107 Reykjavík, Iceland}
\affiliation{Peter Gr\"unberg Institute and Institute for Advanced Simulation, Forschungszentrum J\"ulich and JARA, 52425 J\"ulich, Germany}
\affiliation{Department of Physics, RWTH Aachen University, 52056 Aachen, Germany}

\author{Nikolai~S.~Kiselev}
 \affiliation{Peter Gr\"unberg Institute and Institute for Advanced Simulation, Forschungszentrum J\"ulich and JARA, 52425 J\"ulich, Germany}

\author{Pavel~F.~Bessarab}
\affiliation{Science Institute, University of Iceland, 107 Reykjavík, Iceland}
\affiliation{Department of Physics and Electrical Engineering, Linnaeus University, SE-39231 Kalmar, Sweden}

\date{\today}

\begin{abstract}
We report tailed skyrmions -- a new class of stable soliton solutions of the 2D chiral magnet model. 
Tailed skyrmions have elongated shapes and emerge in a narrow range of fields near the transition between the spin spirals and the saturated state.
We analyze the stability range of these solutions in terms of external magnetic field and magnetocrystalline anisotropy. 
Minimum energy paths and the homotopies (continuous transitions) between tailed skyrmions of the same topological charge have been calculated using the geodesic nudged elastic bands method.
The discovery of tailed skyrmions extends the diversity of already-known solutions illustrated by complex morphology solitons, such as tailed skyrmion bags with and without chiral kinks.
\end{abstract}
\maketitle 

\section{Introduction}

Chiral magnets represent a unique class of materials where the competition between local interaction, in particular, Heisenberg exchange and Dzyaloshinskii-Moriya interaction~\cite{Dzyaloshinskii, Moriya} (DMI), give rise to a vast variety of topological magnetic solitons such as $k\pi$-skyrmions~\cite{Bogdanov_89, Bogdanov_1994, Bogdanov_1994JMMM, Bogdanov_99}, skyrmion bags~\cite{Rybakov_19, Foster_19}, and skyrmions with chiral kinks~\cite{Barton-Singer_20, Kuchkin_20ii}. 
Note that the physical systems that exhibit such diversity of solitons possessing arbitrary topological charges and morphology are very rare in nature. 

The primary parameter that distinguishes magnetic solitons is the topological charge, which defines the homotopy class of a particular solution:
\begin{equation}
Q = \dfrac{1}{4\pi}\int\!  \mathbf{n}\cdot \left(\partial_\mathrm{x}\mathbf{n}\times\partial_\mathrm{y}\mathbf{n}\right) \,\mathrm{d}x\mathrm{d}y,
\label{Qint}
\end{equation}
where $\mathbf{n}$ is the magnetization unit vector field.

The solutions with identical $Q$ can be continuously (without the appearance of magnetic singularities) transformed into each other ~\cite{Rybakov_19}. 
We refer to such transformations as \textit{homotopies}, which can be defined as follows. 
For two magnetic textures $\mathbf{n}_{1}(\mathbf{r})$ and $\mathbf{n}_{2}(\mathbf{r})$ with identical $Q$, there is a unity parametric vector field $\mathbf{n}(\mathbf{r}; \tau)$, $\tau\in[0,1]$, which is continuously differentiable with respect of $\mathbf{r}$ and $\tau$ and obeys the boundary conditions $\mathbf{n}(\mathbf{r};0)=\mathbf{n}_{1}(\mathbf{r})$, $\mathbf{n}(\mathbf{r};1)=\mathbf{n}_{1}(\mathbf{r})$.
Homotopy between any two configurations $\mathbf{n}_{1}(\mathbf{r})$ and $\mathbf{n}_{2}(\mathbf{r})$ is not defined uniquely as there are infinitely many ways to introduce the parameter $\tau$. A constraint can be introduced by considering only those homotopies that represent minimum energy paths (MEPs), i.e. the paths lying lowermost on the energy surface of the system. Homotopies  satisfying the MEP condition are also physically relevant as they define thermal stability of the magnetic configurations within harmonic rate theories. 
For solitons with different $Q$, homotopies are, by definition, impossible.

\begin{figure*}[ht!]
\centering
\includegraphics[width=16.3cm]{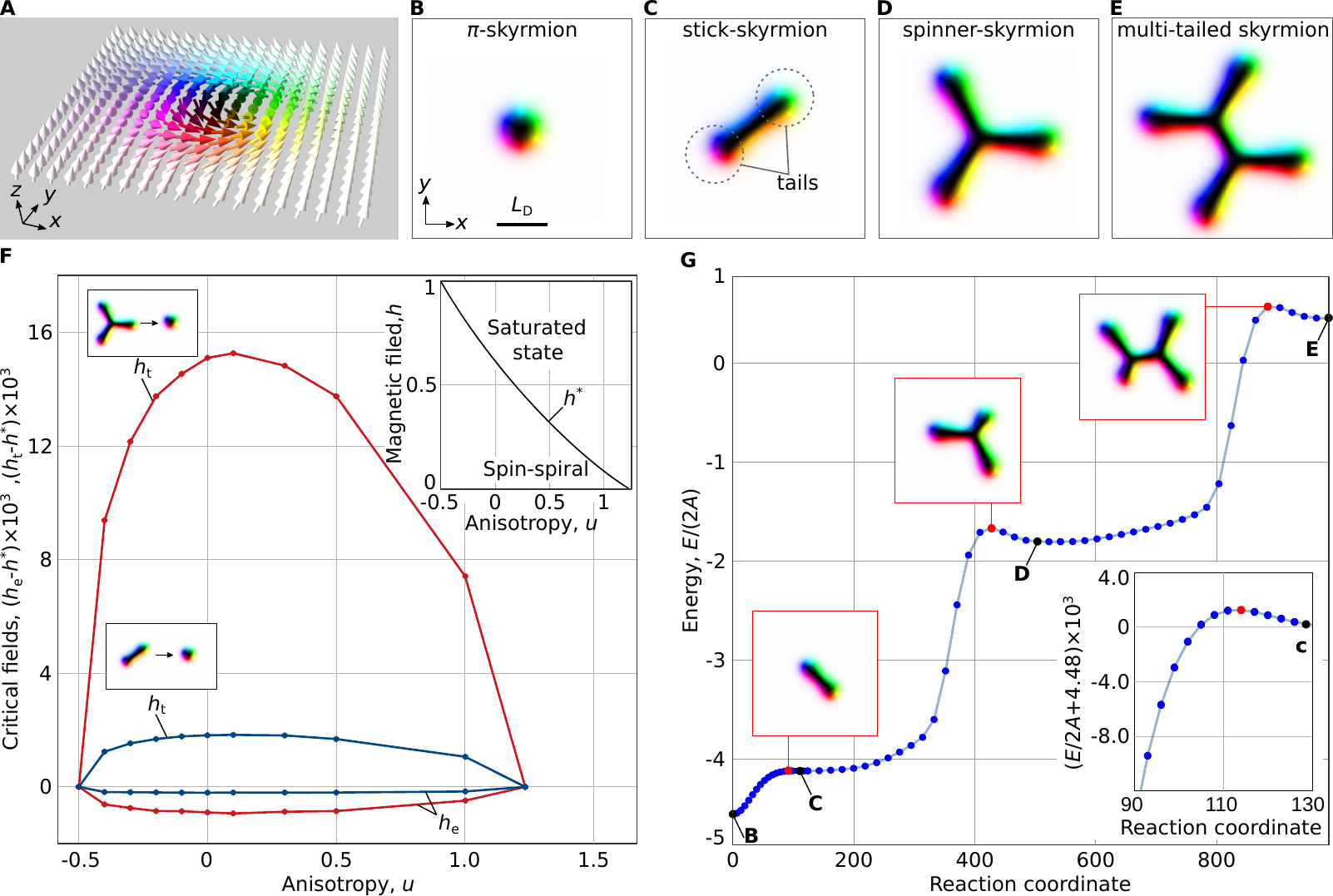}
\caption{~\small \textbf{Tailed skyrmions}.
(A) The magnetization vector field of the $\pi$-skyrmion.
(B) Top view of a $\pi$-skyrmion, where colors encode the magnetization direction.
(C)-(E) Color-coded spin texture of tailed skyrmions with $Q=-1$.
(F) Stability diagram of tailed skyrmions in terms of the magnetic field, $h$, and magnetocrystalline anisotropy, $u$.
(G) Minimum energy path between skyrmions depicted in (B)-(E).
All simulations were performed on the $4L_\mathrm{D}\times 4L_\mathrm{D}$ domain.}
\label{Fig1}
\end{figure*}

The representative and the most well-studied soliton in chiral magnets -- $\pi$-skyrmion with $Q=-1$ -- is shown in Figure~\ref{Fig1}A-B.
$\pi$-skyrmions were theoretically predicted more than thirty years ago by Bogdanov and Yablonskii~\cite{Bogdanov_89, Bogdanov_1994, Bogdanov_1994JMMM}.
Nowadays, there is a long list of magnetic materials where $\pi$-skyrmions were observed experimentally~\cite{Tokura_21}.

Another type of axially symmetric solitons called $k\pi$-skyrmions ($k>1$) was presented by Bogdanov and Hubert~\cite{Bogdanov_99}. 
Despite the morphological diversity of $k\pi$-skyrmions, they all belong to only two homotopy classes, with $Q=0$ and $Q=-1$ for even and odd $k$, respectively.

The non-axially symmetric solutions with arbitrary topological charge, also known as skyrmion bags~\cite{Rybakov_19, Foster_19} and skyrmions with chiral kinks~\cite{Barton-Singer_20, Kuchkin_20ii} were found only recently.
The skyrmion bags with positive topological charge have recently been observed in FeGe plates using transmission electron microscopy~\cite{Tang_21}.
Co-existing skyrmion and its antiparticle (antiskyrmion) i.e. soliton with two chiral kinks were observed experimentally in FeGe thin platelet~\cite{Zheng_22}. 

In this article, we introduce a new class of chiral magnetic skyrmions we refer to as tailed skyrmions. We demonstrate rich diversity of tailed skyrmions and present a comprehensive study of their static properties, stability and finite-temperature dynamics. 
Tailed skyrmions can be thought of as various elongated skyrmions stabilized in the narrow range of external magnetic fields close to the elliptical instability field of an ordinary $\pi$-skyrmion. 
Some examples of tailed skyrmions are shown in Fig.~\ref{Fig1}. 
To our knowledge, this type of solitons has not been reported in the literature. 

\section{Model}

We consider the two-dimensional (2D) micromagnetic model of a chiral magnet containing three energy terms:
\begin{equation}
\mathcal{E}=\int \{w_\mathrm{ex}(\textbf{n}) + w_\mathrm{D}(\textbf{n})+w_\mathrm{U}(\textbf{n})\}\,  l\mathrm{d}x\mathrm{d}y,
\label{Etot}
\end{equation}
where $\textbf{n}=\textbf{M}/M_\mathrm{s}$ is the magnetization unit vector field which is uniform across the film thickness $l$,  $M_\mathrm{s}$ is the saturation magnetization, $w_\mathrm{ex}(\textbf{n})=\mathcal{A}\left|\nabla\mathbf{n}\right|^{2}$ is the Heisenberg exchange interaction and 
$w_\mathrm{U}(\textbf{n})=M_\mathrm{s}B_\mathrm{ext}\left(1-n_\mathrm{z}\right)+\mathcal{K} \left(1-n_\mathrm{z}^{2}\right)$ is the potential term containing the Zeeman interaction and the easy-axis/easy-plane anisotropy. 
The external magnetic field is perpendicular to the plane of the film, $\textbf{B}_\mathrm{ext}\parallel\textbf{e}_\mathrm{z}$.

The DMI term $w_\mathrm{D}(\mathbf{n}) = \mathcal{D} w(\mathbf{n})$ 
is defined by combinations of Lifshitz invariants~\cite{Bogdanov_89}, $\Lambda_{ij}^{(k)}\!=\! n_i\partial_k n_j\!-\!n_j\partial_k n_i$.
The results presented below are valid for systems with
Bloch-type modulations~\cite{ElisaGiovanni}, 
N\'{e}el-type modulations~\cite{Romming_13,Kez_15,Romming_15} as well as for crystals with D$_{2\mathrm{d}}$ or S$_4$ point group symmetry~\cite{Bogdanov_89}. 
Without loss of  generality, we assume Bloch-type DMI, $w(\mathbf{n})\! =  \!\Lambda_{\mathrm{zy}}^{(\mathrm{x})}\!+\! \Lambda_{\mathrm{xz}}^{(\mathrm{y})}\!$, in our calculations.
By introducing the characteristic size of chiral modulations $L_\mathrm{D}=4\pi\mathcal{A}/\mathcal{D}$ and the characteristic magnetic field $B_\mathrm{D}=\mathcal{D}^2/(2 M_\mathrm{s}\mathcal{A})$, we  
reduce the number of independent parameters to two, namely the dimensionless external
magnetic field $h=B_\mathrm{ext}/B_\mathrm{D}$ and anisotropy parameter $u = \mathcal{K}/(M_\mathrm{s} B_\mathrm{D})$.

\section{Methods}

This work involves several numerical methods. 
Stable magnetic configurations are obtained via direct energy minimization using the conjugate gradient method with stereographic projections of the $\mathbf{n}$-field. For details, see Supplementary materials in Ref.~\cite{Rybakov_15}. 
MEPs) between the stable states are calculated using the geodesic nudged elastic band (GNEB) method~\cite{bessarab_2015, bessarab_2017}.
Dynamical properties of tailed skyrmions are investigated via the numerical integration of the Landau-Lifshitz-Gilbert (LLG) equation~\cite{Landau_Lifshitz} using the well-established semi-implicit method from Ref.~\cite{Mentink_10}.

The calculations were performed and double-checked with various software.
In particular, we used Excalibur~\cite{Excalibur} and Mumax3~\cite{Mumax} for  energy minimization and calculation of stability ranges.
MEPs were calculated using the Spirit code~\cite{Spirit}. 
We used high-accuracy finite difference schemes in the energy minimization to remove numerical artifacts in the calculation of derivatives in Eq.~\eqref{Etot}.
A typical mesh density in our simulations is 64 nodes per $L_\mathrm{D}$.
The details can be found in Refs.~\cite{Rybakov_19, Kuchkin_20ii}.

For the MEP calculations and stochastic LLG simulations, we used an effective spin-lattice model of nearest neighbors.
All calculations were performed with periodic boundary conditions in the film plane.
For visualization of the magnetization, we use the standard color code as explained in Figure~\ref{Fig1}A and B. 

\section{Initial guesses}

To obtain statically stable solutions for tailed skyrmions in Mumax3, we constructed initial guesses for the $\mathbf{n}$-field
by placing domains with magnetization antiparallel to the field, $\mathbf{n}=(0,0,-1)$, into the state with magnetization pointing along the field, $\mathbf{n}=(0,0,1)$.
In the case of axially symmetric skyrmions, e.g., $\pi$-skyrmion in Figure~\ref{Fig1}A and B, we use a circular shape domain with magnetization down.
For tailed skyrmions shown in Figure~\ref{Fig1}C-E, the domains with down magnetization should approximately mimic the shape of the skyrmion, which can be achieved by placing circular domains in a chain.
For details, see the Supplementary materials with the corresponding Mumax script.

In the simulation with Excalibur and Spirit, we used the built-in options allowing one to drag and invert the magnetization field interactively, as illustrated in Supplementary Movie 1.
Notably, the interactive modification of the magnetization field with the drag tool permits for examining whether the skyrmion has reached the optimal shape and size. 

\section{Results}

\subsection{Stability diagram}
Figure~\ref{Fig1}C, D, and E show elementary tailed skyrmions.
The two solutions with two and three tails can be referred to as a stick-skyrmion and a spinner-skyrmion, respectively.
We use these representative solutions to illustrate the range of tailed skyrmion existence.
In Figure~\ref{Fig1}F, the critical fields for stick-skyrmion and spinner-skyrmion are shown with blue and red curves, respectively.
The field range corresponding to stable tailed skyrmions is bound from above by the transition field into axially symmetric $\pi$-skyrmion, $h_\mathrm{t}$, and from below by elliptic instability field, $h_\mathrm{e}$.
%
Note that Figure~\ref{Fig1}F shows the deviation of $h_\mathrm{t}$ and $h_\mathrm{e}$ from $h^*$ -- 
the critical field corresponding to the transition between the spin-spiral state and the ferromagnetic phase -- so as to emphasize the difference between the critical fields which are quite close to each other. 
The dependence $h^*(u)$ is the solution of the well-known equation~\cite{Bogdanov_1994JMMM}:
\begin{align}
2\pi-4\sqrt{h+2u}-\dfrac{\sqrt{2}h}{\sqrt{u}}\ln\dfrac{\sqrt{h+2u}+\sqrt{2u}}{\sqrt{h+2u}-\sqrt{2u}}=0,
\label{DWenergy}
\end{align}
which is derived from the criterion that the energy of the isolated spiral ($2\pi$-domain wall) equals the energy of the saturated ferromagnetic state. 
The functional dependence $h^*(u)$ is shown in the inset of Figure~\ref{Fig1}F.  

The critical fields for any tailed skyrmion solutions always meet at two points: i) the Bogomolnyi point $h=1$, $u=-0.5$ and ii) the phase transition point between the spin spiral and ferromagnetic phases $h=0$, $u=\pi^2/8$.
Thus, the stability range of tailed skyrmions is quite significant and includes both in-plane and strong out-of-plane anisotropy systems.

The range of the external magnetic field corresponding to stable spinner-skyrmion (see the domain bound by the red lines in Fig.~\ref{Fig1}F) is roughly eight times larger than that for the stick-skyrmion (see the domain bound by the blue lines).
The stability range for most of the other tailed-skyrmion solutions considered in this work was found to coincide with high precision with the stability range of the spinner-skyrmion.
Therefore, the area bound by the red curves in Figure~\ref{Fig1}F can be considered as a good estimate for the stability range of all tailed-skyrmions.

\subsection{Homotopies and minimum energy paths}

Figure~\ref{Fig1}G shows the MEP connecting the $\pi$-skyrmion, stick-skyrmion, spinner-skyrmion and  four-tailed skyrmion 
(see Figure~\ref{Fig1}B, C, D, and E).
The calculated MEP turns out to be a homotopy as the topological charge of the system remains the same, $Q=-1$, at every point of the MEP. 
Some other examples of homotopy MEPs have already been presented in Ref.~\cite{Kuchkin_22}, reporting the method for finding exotic three-dimensional hybrid skyrmion tubes. 

Starting from the $\pi$-skyrmion solution and following the  MEP we observe a  sequential increase in the number of tails in the magnetic texture. 
The growth of a new tail requires overcoming an energy barrier that turns out to be significantly larger than that for the reverse process, i.e., removal of the tail. The latter defines the stability of the tailed states.
With an increasing external magnetic field, the barriers corresponding to the tail contraction decrease and eventually disappear, leading to the transition of the tailed skyrmions into the axially symmetric $\pi$-skyrmion state.

As follows from the MEP calculations, the energy of the tailed skyrmions increases with the number of tails, meaning that tailed skyrmions represent higher energy metastable states.
The latter is consistent with the previous study~\cite{Melcher_2014} where it was proven that the axially symmetric $\pi$-skyrmion solution represents the global minimum among all solutions with $Q=-1$.

The question about the second lowest energy soliton with $Q=-1$ remains open.
For instance, $3\pi$-skyrmion and some of the skyrmion bags seem to be promising candidates. 
However, comparing the energies of tailed skyrmions, we found that the stick skyrmion may certainly compete with these solutions.
In particular, at $h=0.617$ and $u=0$, the energy of stick-skyrmion is only $8.4\%$ higher than that of the $\pi$-skyrmion, while the energy of $3\pi$-skyrmion for these parameters is $11\%$ higher.
Moreover, the $3\pi$-skyrmion is significantly bigger than the $\pi$-skyrmion and has a diameter of $\sim8L_\mathrm{D}$. The size of other skyrmion bags with $Q=-1$ is even larger.
This means that the stick-skyrmion is not only lower in energy than other solutions but also has the shortest distance to the $\pi$-skyrmion in the configuration space.

\subsection{Thermally activated transition from $\pi$-skyrmion to stick-skyrmion}

\begin{figure}[ht]
\centering
\includegraphics[width=8cm]{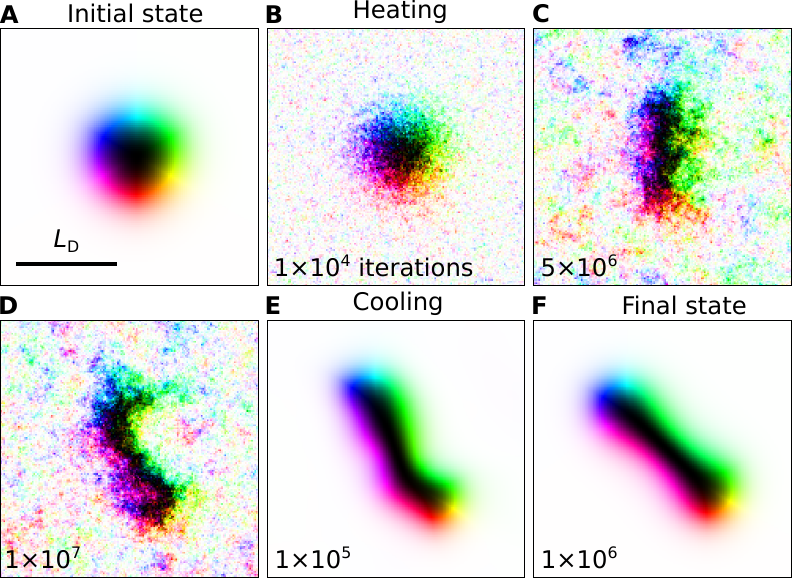}
\caption{~\small Snapshots of the system during the heating-cooling protocol of stochastic LLG simulations. (A) The initial configuration of $\pi$-skyrmion.
(B)-(D) Snapshots of the system after $10^4$, $5\cdot10^{6}$ and $10^{7}$ iterations of stochastic LLG simulations with fixed temperature of $T=0.2J/k_\mathrm{B}$. 
(E), (F) Snapshots after $10^{5}$ and $10^{6}$ LLG time steps with $T=0$. 
Simulations are performed on the $8L_\mathrm{D}\times 8L_\mathrm{D}$ computational domain for $h=0.617$, $u=0$.
}
\label{Fig2}
\end{figure}

Since the tailed stick-skyrmion solution is the closest local minimum to the $\pi$-skyrmion state, excitation of a $\pi$-skyrmion by external stimuli such as thermal fluctuations can induce a nucleation of the stick-skyrmion.
To demonstrate this, we performed finite-temperature spin dynamics simulations for $h=0.617$ and $u=0$. 
We used the LLG time step $\Delta t=0.01J\gamma \mu_\mathrm{s}^{-1}$, and damping parameter $\alpha=0.1$.
The system was simulated at the temperature of $T=0.2J/k_\mathrm{B}$, which is significantly lower than the Curie temperature, which was estimated to be $T_\mathrm{c}=0.7J/k_\mathrm{B}$ for the given system~\cite{Kuchkin_22ii}.

The initial magnetic texture of axially symmetric $\pi$-skyrmion is shown in Figure~\ref{Fig2}A. 
The thermal fluctuations result in skyrmion deformation, as seen in Figure~\ref{Fig2}B-D. 
After turning off the temperature and performing cooling with the LLG simulation at $T=0$, the deformed skyrmion quickly turned into a stick-skyrmion, as shown in Figure~\ref{Fig2}E-F and in Supplementary Movie 2.  
A longer heating of the system, in principle, can also give rise to the appearance of spinner-skyrmion and other multi-tailed skyrmions.
However, such events have a much lower probability because of the large distance between corresponding minima in the parameter space and higher energy barriers, Figure~\ref{Fig1}G.
Nevertheless, our results suggest a simple approach to the experimental observation of tailed skyrmions.

\subsection{Tailed skyrmions with arbitrary topological charge}
So far, we have only discussed solutions with $Q=-1$.
However, the tails can in principle be added to any soliton with an arbitrary topological charge.
Adding a tail creates a new soliton, but does not change the topological charge of the system.%
This section provides examples of such skyrmions.
All solutions presented in the following were obtained on a square simulation domain of size $L_\mathrm{x}=L_\mathrm{y}=8L_\mathrm{D}$ for parameters $h = 0.617$ and $u = 0$.

\begin{figure}[ht]
\centering
\includegraphics[width=8cm]{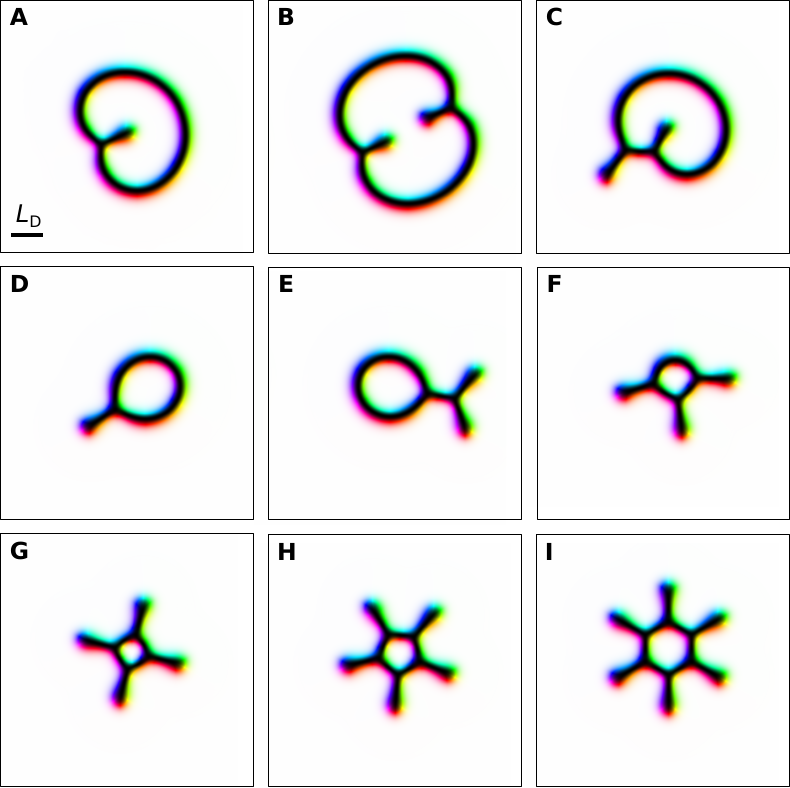}
\caption{~\small Tailed skyrmions with $Q=0$.
Skyrmionium with tails on the inner contour only (A), (B), on both contours (C), and on the outer contour only (D) - (I).
The scale bar is the same for all images.
}
\label{Fig3}
\end{figure}

In Figure~\ref{Fig3}, we show topologically trivial solitons -- skyrmionium with tails.
Note that all solitons shown in Figure~\ref{Fig3} are homotopically equivalent despite different number of tails and their position on the inner or outer contour of the solitons.
Noticeably, the skyrmionium with more than three tails on the outer contour, see, e.g., Figure~\ref{Fig3}G-I, has rotational symmetry of order equal to the number of tails.

\begin{figure}[ht]
\centering
\includegraphics[width=8cm]{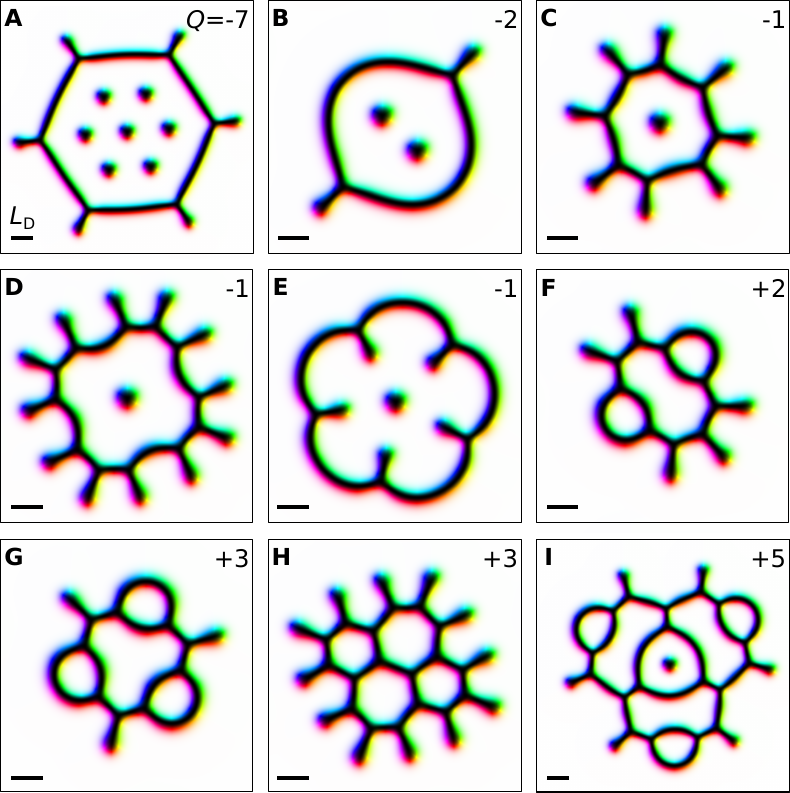}
\caption{~\small Skyrmion bags with tails stabilized for $u=0$ and $h=0.617$.
}
\label{Fig4}
\end{figure}

Figure~\ref{Fig4} provides examples of tailed skyrmion bags of various topological charges and demonstrates diversity of tailed skyrmions.
For example, Figure~\ref{Fig4}C-E shows three skyrmion configurations with the same topological charge $Q=-1$ but different symmetry and different number of tails.
Figure~\ref{Fig4}G and H demonstrates two different skyrmion bags with the same positive topological charge, $Q=+3$.

\begin{figure}[ht]
\centering
\includegraphics[width=8cm]{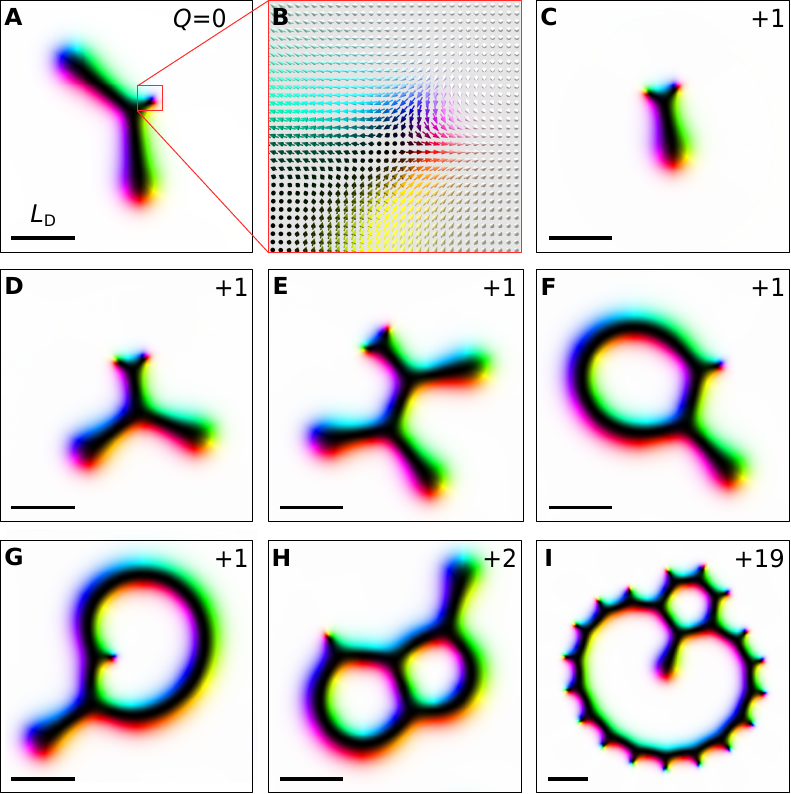}
\caption{~\small Skyrmions with  tails and chiral kinks stabilized for $h=0.617$ and $u=0$.
(B) Detailed view of the spin texture near the chiral kink of skyrmion shown in (A).  
}
\label{Fig5}
\end{figure}

\begin{figure}[ht]
\centering
\includegraphics[width=8.0cm]{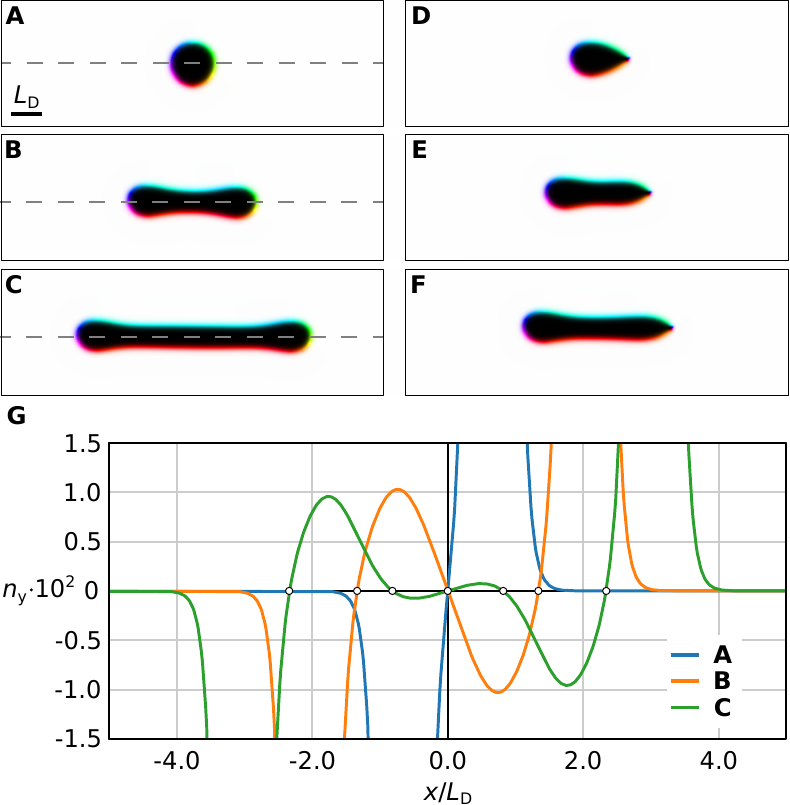}
\caption{~\small 
(A) axially symmetric $\pi$-skyrmion.
(B), (C) stick-skyrmions of different lengths.
(D) chiral droplet soliton.
(E) and (F) stick-skyrmion with one tail and one chiral kink. 
All solutions co-exist for $u=1.194$ and $h=0.01$.
The energy minimization was performed on a large size domain, $L_\mathrm{y}=4L_\mathrm{D}$ and $L_\mathrm{x}= 16L_\mathrm{D}$, to avoid the influence of periodic boundary conditions.
All images have identical scale, as marked in (A).
(G) $y$-component of magnetization along the dashed symmetry line of the  skyrmions (A)-(C).
Hollow circles denote the zeros of $n_\mathrm{y}(x)$.
}
\label{Fig6}
\end{figure}

Another qualitatively new feature appears in the case of skyrmions with chiral kinks and tails.
Figure~\ref{Fig5} provides several examples of such skyrmions having various topological charges. 
Note that the chiral kink on the outer contour of the skyrmion carries the topological charge $Q=+1$.
Because of this, the topological charge of skyrmions in Figure~\ref{Fig5} A and B equals zero and does not depend on the number of tails.  
The skyrmions with antikinks of topological charge $Q=-1$ are only stable at strong out-of-plane anisotropy~\cite{Kuchkin_20ii}.

Figures~\ref{Fig5} C-E illustrate that the chiral kinks form coupled pairs in the presence of tails.
A similar phenomenon was observed for chiral kinks on the $2\pi$-domain wall~\cite{Kuchkin_20ii}.
Generally, the chiral kinks on a straight domain wall repel each other at large distances and attract each other only at small distances~\cite{Kuchkin_20ii}.
Similarly, a pair of kinks can form a coupled state on tailed skyrmions, with finite distance between the kinks on the order of $0.1L_D$. 
With increasing field, the solutions with $Q=+1$ shown in Figures~\ref{Fig5} C-E  collapse to an antiskyrmion~\cite{Kuchkin_20i}. 
Figures~\ref{Fig5} F-I show skyrmion bags with positive topological charges hosting chiral kinks and tails simultaneously.

\subsection{Effect of strong perpendicular anisotropy on tailed skyrmions}

So far, the diversity of tailed skyrmions has been demonstrated for the $u=0$ case.
However, all tailed skyrmion solutions can also be obtained for any value of $u$ in the range from $-1/2$ to $\pi^2/8$ (see the diagram in Fig.~\ref{Fig1}F).  
Moreover, for strong perpendicular anisotropy, $u\lesssim \pi^{2}/8$, we found distinct co-existing solutions representing tailed skyrmions of different lengths.
In Figure~\ref{Fig6}B, C, we show several examples of stick-skyrmions of different lengths, which coexist for a given $u$, $h$, and are separated by small energy barriers.
The energy of the stick-skyrmion increases with its length.
Note that such tailed skyrmions of different lengths are observed only at strong anisotropy. 
As Figure~\ref{Fig6}G shows, the magnetic structure of such stick-skyrmions is non-trivial. In particular, the $y$ component of magnetization is modulated along the skyrmion symmetry axis giving an energy gain sufficient for the skyrmion stabilization. 
The number of zeros of $n_\mathrm{y}$ increases with the length of a stick-skyrmion.

Interestingly, the stick-skyrmion with one tail and one chiral kink (see Figure~\ref{Fig6}D-F) also exhibits multiple minima corresponding to the solutions of different lengths.

\section{Conclusions}

In conclusion, we presented in this work a new class of chiral skyrmions -- tailed skyrmions. 
We identified the range of uniaxial anisotropy and perpendicular magnetic field where the tailed skyrmions remain statically stable.
We show a wide diversity of tailed skyrmions with arbitrary topological charges.
The presence of tails changes only the soliton's symmetry but does not change its topological charge.
Thus, the transitions between tailed and tail-free solutions are homotopies.
We demonstrated that such transitions can be identified using the GNEB method. 
The energy barriers involved in the homotopies tend to be smaller than what needs to be overcome to change the topological charge of the system. 
As a result, tailed skyrmions can be obtained from a regular $\pi$-skyrmion by applying weak thermal fluctuations, as demonstrated by the stochastic LLG simulations.

Our results contribute to the development of a complete picture of the diversity of statically stable magnetic solitons in chiral magnets.

\begin{center}
    {\footnotesize\bf{ACKNOWLEDGMENTS}}
\end{center}
We are grateful to F. Rybakov for fruitful discussions and providing us with the Excalibur code.
The authors acknowledge financial support from the Icelandic Research Fund (Grant No. 217750), the University of Iceland Research Fund (Grant No. 15673), the Swedish Re-
search Council (Grant No. 2020-05110), and the European Research Council (ERC) under the European Union's Horizon 2020 research and innovation program (Grant No.\ 856538, project ``3D MAGiC'').

\end{document}